ARTICLE  OPEN

# Large-signal model of 2DFETs: compact modeling of terminal charges and intrinsic capacitances

Francisco Pasadas [1]*, Enrique G. Marin [2,3], Alejandro Toral-Lopez [2,4], Francisco G. Ruiz [2], Andrés Godoy [2,4], Saungeun Park [5], Deji Akinwande [5] and David Jiménez [1]

We present a physics-based circuit-compatible model for double-gated two-dimensional semiconductor-based field-effect transistors, which provides explicit expressions for the drain current, terminal charges, and intrinsic capacitances. The drain current model is based on the drift-diffusion mechanism for the carrier transport and considers Fermi–Dirac statistics coupled with an appropriate field-effect approach. The terminal charge and intrinsic capacitance models are calculated adopting a Ward–Dutton linear charge partition scheme that guarantees charge conservation. It has been implemented in Verilog-A to make it compatible with standard circuit simulators. In order to benchmark the proposed modeling framework we also present experimental DC and high-frequency measurements of a purposely fabricated monolayer $MoS_2$-FET showing excellent agreement between the model and the experiment and thus demonstrating the capabilities of the combined approach to predict the performance of 2DFETs.

npj 2D Materials and Applications (2019)3:47 ; https://doi.org/10.1038/s41699-019-0130-6

## INTRODUCTION

Since the emergence of graphene, over a surprisingly short period of time, an entire new family of two-dimensional materials (2DMs) has been discovered.[1] Some of them are already used as channel materials in FETs, being promising candidates to augment silicon and III–V compound semiconductors via heterogeneous integration for advanced applications. In this context, the development of electrical models for 2D semiconductor-based FETs (2DFETs) is essential to: (1) interpret experimental results; (2) evaluate their expected digital/RF performance; (3) benchmark against existing technologies; and (4) eventually provide guidance for circuit design and circuit-level simulations.

Several compact models for three-terminal FETs based on graphene and related 2D materials (GRMs) have been recently published encompassing both static and dynamic regimes.[2–10] In particular, Suryavanshi et al.[2] proposed a semi-classical transport approach for the drain current combining the intrinsic FET behavior with models of the contact resistance, traps and impurities, quantum capacitance, fringing fields, high-field velocity saturation and self-heating. However, the dynamic regime is described using a charge model derived for bulk MOSFETs[11] that fails to capture the specific physics of the 2D channel, since it considers that the channel is always in a weak-inversion regime and therefore the channel charge can be assumed to be linearly distributed along the device. Wang et al.[3] reported on a compact model restricted to the static regime of graphene FETs and based on the "virtual-source" approach, valid for both the saturation and linear regions of device operation. Also based on the "virtual-source" approach, Rakheja et al.[4] studied quasi-ballistic graphene FETs. Nevertheless, they also employed the approximation derived for bulk MOSFETs[11] for the dynamic regime description even though the channel material considered is graphene. Jiménez early developed a model describing the drain current for single layer transition metal dichalcogenide (TMD) FETs based on surface potential and drift-diffusion approaches but the dynamic behavior of the TMD-FETs was not treated.[5] Also based on surface potential and drift-diffusion arguments, Parrish et al.[6] presented a compact model for graphene-based FETs for linear and non-linear circuits, but the dynamic regime description was roughly estimated by equally splitting the total gate capacitance between gate-drain and gate-source capacitances. A charge-based compact model for TMD-based FETs only valid for the static regime but simultaneously including interface traps, ambipolar current behavior, and negative capacitance was proposed by Yadav et al.[7] Taur et al.[8] presented an analytic drain current model for short-channel 2DFETs combining a subthreshold current model based on the solution to 2D Poisson's equation and a drift-diffusion approach to get an all-region static description. Rahman et al.[9] discussed a physics-based compact drain current model for monolayer TMD-FETs considering drift-diffusion transport description and the gradual channel approximation to analytically solve the Poisson's equation. Unfortunately, none of them faced the dynamic regime operation. Finally, Gholipour et al.[10] proposed a compact model that combines both the drift-diffusion and the Landauer–Buttiker approaches to properly describe the long- and short-channel FETs in flexible electronics, respectively. As a key feature, the mechanical bending is considered by linearly decreasing the channel material bandgap with respect to the applied strain. However, the capacitance model was roughly approximated by considering that the top (back) gate-drain and top (back) gate-source capacitances are just the geometrical top (back) gate capacitance. In summary, up to date the dynamic description of 2DFETs has been either roughly approximated or not specifically addressed for such devices. Our aim is to develop an accurate and physics-based compact model of the terminal charges and corresponding intrinsic capacitances of four-terminal (4T) 2DFETs valid for all operation regimes. In doing so, we present such a model together with the drain current equation, developed by

[1]Departament d'Enginyeria Electrònica, Universitat Autònoma de Barcelona, 08193 Bellaterra (Cerdanyola del Vallès), Spain. [2]Departamento de Electrónica y Tecnología de Computadores, Universidad de Granada, 18071 Granada, Spain. [3]Dipartimento di Ingegneria dell'Informazione, Università di Pisa, 56122 Pisa, Italy. [4]Excellence Research Unit "Modeling Nature" (MNat), Universidad de Granada, 18071 Granada, Spain. [5]Department of Electrical and Computer Engineering, The University of Texas, Austin, TX 78758, USA.
*email: francisco.pasadas@uab.es

Published in partnership with FCT NOVA with the support of E-MRS

npj nature partner journals

some of us,[12] based on drift-diffusion theory where the surface potential is the key magnitude of the model. So that, the combination of both approaches gives rise to a compact large-signal model suitable for commercial TCAD tools employed in the simulation of circuits based on 2DFETs. The model has been shown to reproduce with good agreement DC and high-frequency measurements of a fabricated n-type MoS$_2$-based FET, and has been used to predict the output transient of a three-stage ring oscillator.

## RESULTS

### Large-signal model of four-terminal 2DFETs

We first deal with the development of the large-signal model of 4T 2DFETs by the sequential formulation of the electrostatics, the drain current equation, the terminal charge model, and the corresponding intrinsic capacitance description.

The cross-section of the considered dual-gated 2DFET is depicted in Fig. 1a. For the sake of brevity, we derive here the expressions for an n-type device (the extension to a p-type transistor can be obtained straightforwardly). Provided that the semiconductor bandgap is not too small, we can assume that $p(x) \ll n(x)$ in all regimes of operation, where $p(x)$ and $n(x)$ are the hole and electron carrier densities, respectively. Upon application of 1D Gauss' law to the double-gate stack shown in Fig. 1b, the electrostatics can be described using the equivalent circuit depicted in Fig. 1c:

$$Q_{net}(x) + Q_{it}(x) = -C_t(V_g - V_{g0} - V(x) + V_c(x)) - C_b(V_b - V_{b0} - V(x) + V_c(x)), \quad (1)$$

where $Q_{net}(x) = q(p(x) - n(x))$ is the overall net mobile sheet charge density and $q$ is the elementary charge. $C_t = \varepsilon_0 \varepsilon_t / t_t$ ($C_b = \varepsilon_0 \varepsilon_b / t_b$) is the top (back) oxide capacitance per unit area where $\varepsilon_t$ ($\varepsilon_b$) and $t_t$ ($t_b$) are the top (back) gate oxide relative permittivity and thickness, respectively; and $V_g - V_{g0}$ ($V_b - V_{b0}$) is the overdrive top (back) gate voltage. These latter quantities comprise work-function differences between the gates and the 2D channel and any possible additional charges due to impurities or doping. The energy $qV_c(x) = E_C(x) - E_F(x)$ represents the shift of the quasi-Fermi level with respect to the conduction band edge and $-qV(x) = E_F(x)$ is the quasi-Fermi level along the channel. This latter quantity must fulfill the boundary conditions: (1) $V(x=0) = V_s$ (source voltage) at the source end; (2) $V(x=L) = V_d$ (drain voltage) at the drain end.

2D crystals are far from perfect and suffer from diverse non-idealities. In particular, interface traps are unavoidable in FETs, even for those processed by state-of-the-art CMOS technology,[13] impacting negatively the performance of 2DFETs.[14] Therefore, it is mandatory to include them in order to achieve an accurate predictive TCAD tool. For n-type devices, acceptor-like traps (which are negatively charged when occupied by electrons and are energetically located in the upper half of the bandgap[15]) contribute the most to the device electrical characteristics.[16] Assuming the traps are situated at an effective energy $E_{it} = -qV_{it}$ below the conduction band, the occupied trap density can be written as $n_{it} = N_{it}/(1 + \exp(V_c - V_{it}/V_{th}))$, where $N_{it}$ is the effective trap density considered to be a delta function in energy. The trap charge density is then $Q_{it} = -qn_{it}$ and the interface trap capacitance $C_{it}$ at the oxide–semiconductor interface (taken as a combination from both interfaces of the ultra-thin 2D channel) can be computed as:

$$C_{it} = \frac{dQ_{it}}{dV_c} = \frac{qN_{it}}{2V_{th}} \frac{1}{1 + \cosh\left(\frac{V_c - V_{it}}{V_{th}}\right)}, \quad (2)$$

where $V_{th} = k_B T/q$ is the thermal voltage, $k_B$ is the Boltzmann constant, and $T$ is the temperature. The term $C_{it}$ adds to the quantum capacitance $C_q$, as shown in Fig. 1c.

We can, thus, find an expression to calculate the overall net mobile sheet density assuming a parabolic dispersion relationship modeled in the effective mass approximation and using Fermi–Dirac statistics[12]:

$$Q_{net}(x) = -qn(x) = -C_{dq}V_{th}u(V_c); \quad u(V_c) = \ln\left(1 + e^{-\frac{V_c}{V_{th}}}\right), \quad (3)$$

where $C_{dq} = q^2 D_0$ is defined as the degenerated-quantum capacitance, which corresponds to the upper-limit achievable when the 2D carrier density becomes heavily degenerated.[17] $D_0 = g_K(m^K/2\pi\hbar^2) + g_Q(m^Q/2\pi\hbar^2)\exp[-\Delta E_2/k_B T]$ is the 2D density of states, with $\hbar$ being the reduced Planck's constant, $g_K$ ($g_Q$) the degeneracy factor and $m^K$ ($m^Q$) the conduction band effective mass at the $K$ ($Q$) band valley. In most TMDs, the second conduction valley, labeled as $Q$ valley, is non-negligible since the energy separation between the $K$ and $Q$ conduction valleys, $\Delta E_2$, is only around $2k_B T$.[18,19] Thus, two conduction band valleys may participate in the transport process. The rest of valleys are, on the contrary, far away in energy to contribute to the electrical conduction,[20] and hence can be neglected for practical purposes.

The quantum capacitance of the 2DM can be computed by evaluating $C_q = dQ_{net}/dV_c$ resulting in:

$$C_q = C_{dq}(1 - e^{-u}). \quad (4)$$

Due to the complexity of Eq. (1) and considering the relation between $Q_{net}$ and $V_c$ given by Eq. (3), it is not possible to get an explicit expression for $V_c$ as a function of the applied bias. However, an iterative Verilog-A algorithm can be implemented to evaluate the chemical potential at the source and drain edges, which are the relevant quantities determining the drain current. In doing so, $V_{cs} = V_c|_{V=V_s}$ and $V_{cd} = V_c|_{V=V_d}$ are obtained, respectively, from Eqs. (1) and (3) by a construct[21,22] which lets the circuit simulator to iteratively solve both equations during run-time.

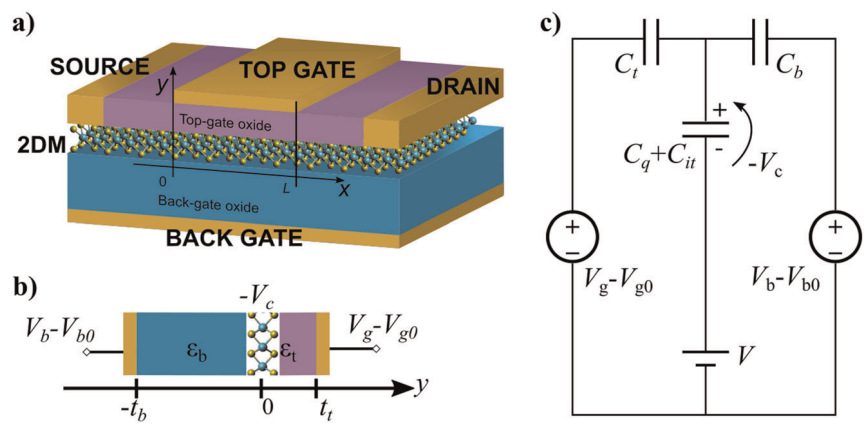

**Fig. 1** **a** Cross-section of the 2DFET considered in the modeling framework. A 2DM plays the role of the active channel. **b** Scheme of the 2DM-based capacitor showing the relevant physical and electrical parameters. **c** Equivalent capacitive circuit of the 2DFET.

In the diffusive regime, the drain current of a 2DFET can be accurately computed by the following compact expression[12]:

$$I_{ds} = \mu \frac{W}{L} C_{dq} V_{th}^2 \left[ \left(1 + \frac{C_{dq}}{C_{tb}}\right) \left(\frac{u_s^2 - u_d^2}{2}\right) + (e^{-u_d} - e^{-u_s}) \right], \quad (5)$$

where $u_s = u(V_{cs})$ and $u_d = u(V_{cd})$; $C_{tb} = C_t + C_b$; $W$ and $L$ are the device width and length, respectively; and $\mu$ is the carrier mobility which has been assumed to be dependent of the applied electric field, carrier density, and temperature as in ref.[2] Once we have a suitable expression for describing the static behavior of the device, the dynamic response must be analyzed to build a large-signal model. In doing so, the terminal currents in the time domain can be computed as:

$$i_k(t) = \frac{dQ_k}{dt} = C_{kg}\frac{dv_g}{dt} + C_{kd}\frac{dv_d}{dt} + C_{ks}\frac{dv_s}{dt} + C_{kb}\frac{dv_b}{dt}, \quad (6)$$

where $k$ stands for g, d, s, and b, i.e. top gate, drain, source and back-gate, respectively. The terminal currents, thus, can be obtained by either computing the time derivative of the corresponding terminal charge or by the weighted sum of the time derivative of the terminal voltages, where the weights are given by the intrinsic capacitances.

The intrinsic capacitances of FETs are modeled in terms of the terminal charges. Specifically, from the electrostatics given in Eq. (1) the following relations are derived[23]:

$$Q_{gb} = Q_g + Q_b = -W \int_0^L Q_{net}(x) dx, \quad (7)$$

$$Q_g = Q_0 - \frac{WC_t}{C_{tb}} \int_0^L Q_{net}(x) dx, \quad (8)$$

$$Q_b = -Q_0 - \frac{WC_b}{C_{tb}} \int_0^L Q_{net}(x) dx, \quad (9)$$

where $Q_0 = WLC_tC_b(V_g - V_{g0} - V_b + V_{b0})/C_{tb}$. The charge controlled by both the drain and source terminals is computed based on the Ward–Dutton's linear charge partition scheme,[24] which guarantees charge conservation:

$$Q_d = W \int_0^L \frac{x}{L} Q_{net}(x) dx, \quad (10)$$

$$Q_s = -(Q_g + Q_b + Q_d) = W \int_0^L \left(1 - \frac{x}{L}\right) Q_{net}(x) dx. \quad (11)$$

The above expressions can be conveniently written using $u$ as the integration variable, as it was done to model the drain current in Eq. (5). Based on the fact that the drain current is the same at any point $x$ along the channel (i.e. we are under the quasi-static approximation framework), we get from the transport model the following equations needed to evaluate the charges in Eqs. (7)–(11):

$$\begin{aligned} dx &= \frac{\mu W}{I_{ds}} Q_{tot}(u) \frac{dV}{dV_c} \frac{dV_c}{du} du \\ x &= \frac{\mu W}{I_{ds}} \left[ \int_{u_s}^{u} Q_{tot}(u) \frac{dV}{dV_c} \frac{dV_c}{du} du \right] \end{aligned}. \quad (12)$$

Replacing Eq. (12) in Eqs. (7) and (10), the following compact expressions are obtained for the terminal charges $Q_{gb}$ and $Q_d$:

$$Q_{gb} = Q_{2D} \frac{2\left(3C_{tb}(e^{u_d}(1+u_s) - e^{u_s}(1+u_d)) + e^{u_s}e^{u_d}\left(C_{dq}+C_{tb}\right)\left(u_s^3 - u_d^3\right)\right)}{6C_{tb}(e^{u_d} - e^{u_s}) + 3e^{u_s}e^{u_d}\left(C_{dq}+C_{tb}\right)\left(u_d^2 - u_s^2\right)},$$
$$Q_d \approx -Q_{2D}\frac{2}{15}\frac{3u_d^5 - 5u_d^3 u_s^2 + 2u_s^5}{\left(u_d^2 - u_s^2\right)^2} \quad (13)$$

where $Q_{2D} = WLC_{dq}V_{th}$. Equation (13) can be replaced in Eqs. (8), (9) and (11) to get the terminal charges $Q_g$, $Q_b$, and $Q_s$, respectively. $Q_d$ in Eq. (13) has been simplified by assuming that the term $e^{(2u_d+2u_s)}$ is prevalent over other low-order terms, e.g., $e^{u_d}$ and $e^{u_s}$.

Once the terminal charges are evaluated, a four-terminal FET can be modeled with 4 self-capacitances and 12 intrinsic transcapacitances given by:

$$C_{ij} = -\frac{\partial Q_i}{\partial V_j} \quad i \neq j \qquad C_{ij} = \frac{\partial Q_i}{\partial V_j} \quad i = j \qquad i,j = g,d,s,b, \quad (14)$$

where $C_{ij}$ describes the dependence of the charge at terminal $i$ with respect to a varying voltage applied to terminal $j$ assuming that the voltage at any other terminal remains constant. Due to charge conservation and considering a reference-independent model, only nine independent capacitances are needed to describe the 4T device. In addition, taking advantage of the relations between the top- and back-gate capacitances,[25] the computation of only four capacitances is enough; for instance, $C_{gg}$, $C_{gd}$, $C_{dg}$, and $C_{dd}$ can be chosen as the independent set.

Model validation

In order to validate the presented model, we perform the DC and RF measurements of an experimental monolayer $MoS_2$-FET. The fabricated device consists of an n-type channel of a chemical vapor deposited (CVD) $MoS_2$ monolayer transferred onto 285 nm intrinsic $Si/SiO_2$ wafer. The 150-nm-long $MoS_2$ channel is contacted with a stack of 2/70 nm Cr/Au and electrostatically controlled by an embedded gate formed by a 10-nm barrier of atomic layer deposited $Al_2O_3$ and a gate metal stack consisting of 2/23 nm Ti/Au. More details on the fabrication and characterization process can be found in the Methods section. Table 1 summarizes the input parameters used for simulating the device under test. The extracted 2D semiconductor–metal contact resistance for the device is 3.5 kΩ µm and has been included in the simulation by connecting lumped resistors to the source and drain terminals.

We first test our model by comparing the DC transfer characteristics (TCs) obtained at two different drain voltages. A good agreement between both, measurements and simulations, has been achieved as shown in Fig. 2. The model predicts the DC behavior accurately in all regimes of operation with current values ranging between $10^{-3}$ up to $10^2$ µA/µm. However, the experimental device departs from the exponential trend for values in the limit of the IRDS specification for low-power applications, being the OFF current actually limited by the onset of hole current at negative $V_{gs}$ (see Supplementary Notes and Supplementary Fig. S1 for a detailed discussion). This small mismatch, however, does not impact the dynamic operation prediction. Indeed, we have assessed the expected RF performance of such device in Fig. 3 by benchmarking the simulated small-signal current gain (h21) and unilateral power gain (U) against the results extracted from the experimental S-parameters (after de-embedding) measured in the range 0.1–15 GHz. Both, simulated $h21$ and $U$, are predicted with excellent agreement to the experimental measurements. In order to evaluate the RF performance, we calculate the intrinsic capacitances, plotted in Fig. 4 (solid lines). The relative error between the numerical solution of Eq. (14) by using the expressions in Eqs. (8)–(11) and the analytical computation of Eq. (14) by using the compact expressions in Eq. (13) is less than 2% for the bias window considered in Fig. 4. The results are presented together with the intrinsic capacitances relying on the well-known conventional charge model developed for bulk MOSFETs[11] and elsewhere used to model 2DFETs[2]. As can be seen, the latter is accurate only either when a low drain bias is applied or if the device is operated in the subthreshold regime (see Supplementary Notes and Supplementary Fig. S2 for more details).

Table 1. Input parameters of the 2DFET under test.

| | | | |
|---|---|---|---|
| $L$ | 150 nm | $\mu_0$ | 21.2 cm$^2$/Vs |
| $W$ | 10 µm | $m^{K\ 26}$ | 0.48$m_0$ |
| $L_t$ | 10 nm | $m^{Q\ 26}$ | 0.57$m_0$ |
| $L_b$ | 285 nm | $m_0$ | 9.11 × 10$^{-31}$ kg |
| $\varepsilon_t$ | 9 | $g_K^{27}$ | 2 |
| $\varepsilon_b$ | 3.9 | $g_Q^{27}$ | 6 |
| $V_{g0}$ | 0.42 V | $\Delta E_2^{18}$ | 0.11 eV |
| $V_{b0}$ | 0 V | $N_{it}$ | 1.8 × 10$^{16}$ m$^{-2}$ |
| $T$ | 300 K | $V_{it}$ | 0.085 eV |

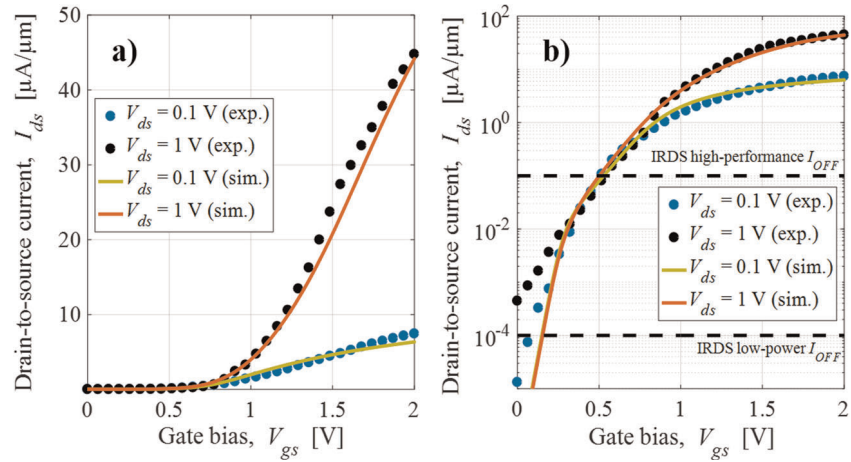

**Fig. 2** Transfer characteristics for different drain voltages in **a** linear and **b** logarithmic scales. Simulations are plotted with solid lines, and experimental measured data with symbols. Dashed lines in **b** highlight both the IRDS high-performance and low-power $I_{OFF}$ current limits.

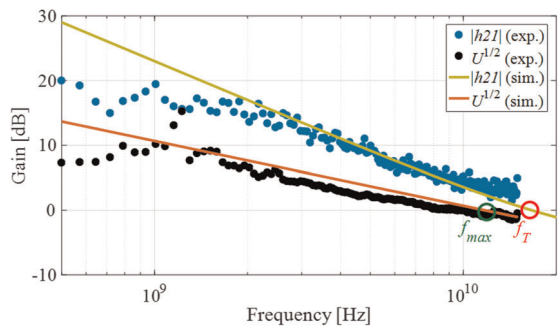

**Fig. 3** Simulated (solid line) and measured (symbols) RF performance of the MoS$_2$-based FET described in Table 1 ($V_{gs}$ = 4.4 V; $V_{ds}$ = 3.5 V). The device shows a cut-off frequency $f_T$ = 20.2 GHz and a maximum oscillation frequency $f_{max}$ = 11.3 GHz.

The trends in the intrinsic capacitances can be better understood looking at the overdrive gate and drain bias dependences of both gate ($Q_g$) and drain ($Q_d$) charges, which are depicted in Fig. 5a, b, respectively. Figure 5c, d shows the same bias dependence of a set of four independent intrinsic capacitances of the 4T 2DFET under test. As the device is an n-type transistor, we have also plotted the shift of the quasi-Fermi level with respect to the conduction band edge at both the drain/source sides (corresponding to the chemical potentials $V_{cd}$ and $V_{cs}$) in Fig. 5e, f. For the sake of clarity, we provide a separate discussion of the intrinsic capacitances according to the transistor operation regime.

*Subthreshold regime (OFF operation)*: If both $V_{cd}$ and $V_{cs}$ > 10$V_{th}$, then $Q_g$, $Q_d$ ~ 0 (see Fig. 5a). Therefore the channel is empty of carriers and the device operates in the OFF state. This situation happens for overdrive gate biases lower than the threshold voltage, $V_A$, according to Fig. 5e. Given the aforementioned conditions then $C_{gg} \approx C_t C_b WL/C_{tb}$ and $C_{gd} \approx C_{dg} \approx C_{dd} \approx 0$, see to the left of A1 point in Fig. 5c.

*Saturation regime (ON operation)*: On the other hand, when $V_{cd}$ > 10$V_{th}$ while $V_{cs}$ < 10$V_{th}$, the pinch-off is originated at the drain side. This situation is produced for overdrive gate biases between $V_A$ and $V_B$ in Fig. 5e ("A–B" section). Both $C_{gg}$ ("A1–B3" section) and $C_{dg}$ ("A1–B2" section) jump because the channel charge at the source side becomes very sensitive to the Fermi level location when it is close to the conduction band edge. However, $C_{gd}$ and $C_{dd}$ are negligible ("A1–B1" section) because the depletion region close to the drain prevents this terminal to control the channel charge. The saturation regime is also observed at the right column of Fig. 5 for drain biases higher than $V_C$ (drain saturation voltage) in Fig. 5f (right of C point). The intrinsic capacitances $C_{gg}$ (right of C3 point) and $C_{dg}$ (right of C2 point), shown in Fig. 5d, decrease with respect to the linear regime (left of C point, explained below) because the channel is depleted at the drain side, so both $Q_g$ and $Q_d$ become less sensitive to $V_{gs}$ as compared to the linear regime. On the other hand, $C_{gd}$ and $C_{dd}$ are negligible (right of C1 point), as $Q_g$ and $Q_d$ cannot be controlled anymore by the drain terminal after the depletion of the channel drain side, which is confirmed in Fig. 5b.

*Linear (ohmic) regime (ON operation)*: This regime takes place when both $V_{cd}$ and $V_{cs}$ < 10$V_{th}$. This situation is produced for gate overdrive voltages higher than $V_B$ in Fig. 5e (right of B point), where the channel starts to leave the depletion scenario at the drain side. Then a jump is observed in $C_{gg}$ (right of B3 point), $C_{dg}$ (right of B2 point), $C_{gd}$ and $C_{dd}$ (right of B1 point) as a consequence of the recovered electrical connection between the channel and the drain terminal (see Fig. 5a, c). The linear regime can be also observed at the right column of Fig. 5 for drain biases lower than $V_C$ (see Fig. 5f). The negative value of $V_{cs}$ means that the channel is degenerated at the source side for the specific bias considered in this study. In Fig. 5d, it can be seen that $C_{gg}$ (left of C3 point), $C_{dg}$ (left of C2 point), $C_{gd}$ and $C_{dd}$ (left of C1 point) decrease with $V_{ds}$ as the channel get closer to the drain depletion condition reached at $V_{ds} = V_C$.

Finally, with the aim of showing the potential of the developed TCAD tool, a relevant circuit commonly used to evaluate the performance of a digital technology, namely, a ring oscillator (RO) is simulated. The design of a three-stage based RO encompasses both DC and transient simulations. Each device has been described by the parameters shown in Table 1 that have been demonstrated to accurately reproduce the fabricated MoS$_2$-based

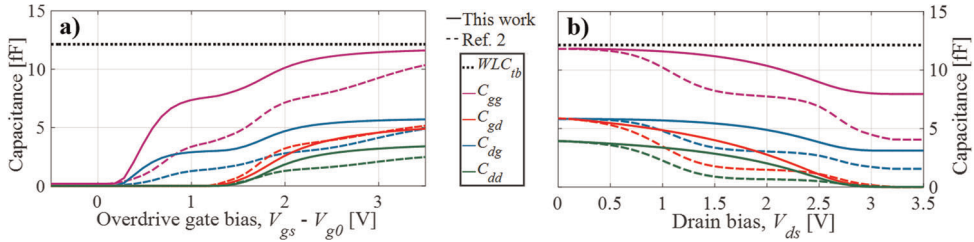

**Fig. 4** Intrinsic capacitances versus **a** overdrive gate bias (at $V_{ds} = 1$ V) and **b** drain bias (at $V_{gs} - V_{g0} = 3.4$ V). Solid lines represent our model outcome and dashed lines show the data calculated using the model adopted in ref. [2]

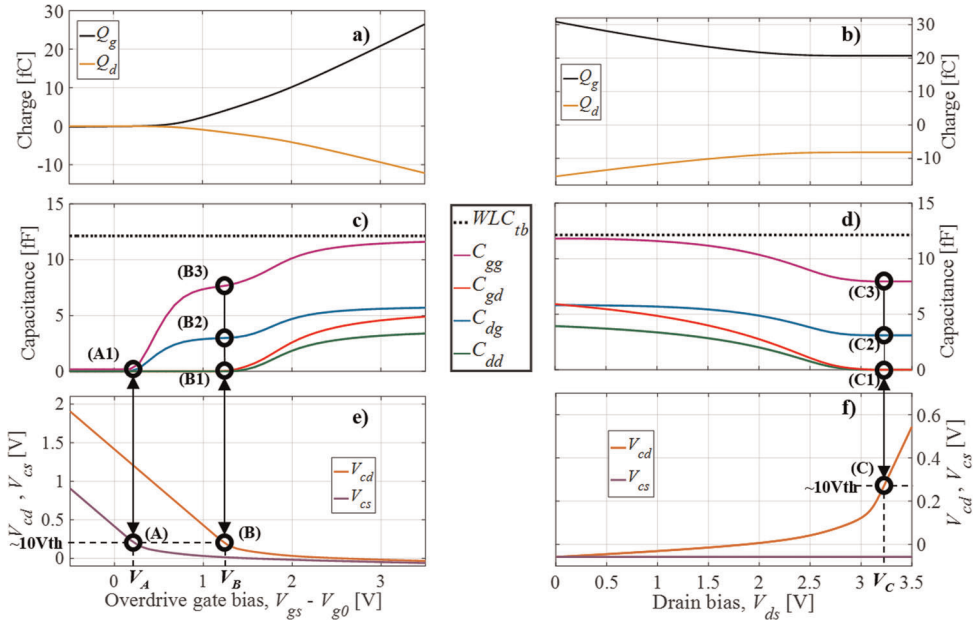

**Fig. 5** Gate and drain charges versus **a** overdrive gate bias (at $V_{ds} = 1$ V) and **b** drain bias (at $V_{gs} - V_{g0} = 3.4$ V). Intrinsic capacitances versus **c** overdrive gate bias (at $V_{ds} = 1$ V) and **d** drain bias (at $V_{gs} - V_{g0} = 3.4$ V). Shift of the Fermi level with respect to the conduction band edge, namely, the chemical potential at the drain ($V_{cd}$) and source ($V_{cs}$) sides versus **e** overdrive gate bias (at $V_{ds} = 1$ V) and **f** drain bias (at $V_{gs} - V_{g0} = 3.4$ V).

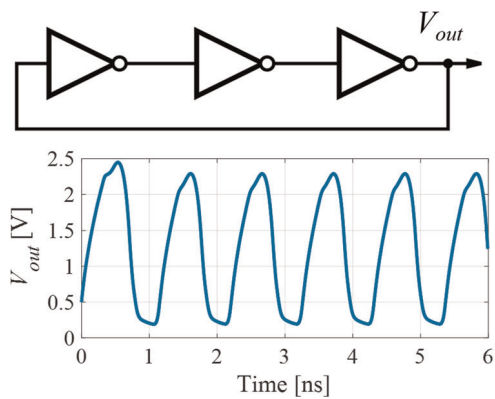

**Fig. 6** Three-stage ring oscillator switching at a frequency around 1 GHz based on the 2DFETs described in Table 1. The supply bias is 3.5 V.

FET in Figs. 2 and 3. Specifically, we have designed the three-stage RO to oscillate at a specific frequency $f_{osc} = 3^{1/2}/(2\pi R_{dd}C) \approx 1$ GHz by using a resistor connected to each drain of $R_{dd} = 20.4$ kΩ and capacitors connected to the output of each stage with $C = 13.5$ fF. In doing so, each single stage has been designed to show a voltage gain of 3.7 at $f_{osc}$, upon biasing it at $V_{gs} = 1.25$ V and $V_{ds} = 3.5$ V. Figure 6 shows how the technology assessed in this work provides a suitable switching characteristic around 1 GHz. The simulation predicts a distorted sinusoidal as the AC component goes over a large range of bias with imperfect linearity. It is worth noticing that a small-signal model (in contrast with a large-signal model) could not predict that feature. Supplementary Fig. S3 provides a comparison of the compact model outcome against experimental measurements of bilayer $MoS_2$ FETs[26] comprising the assessment of transfer characteristics, output characteristics, the performance of an inverter, and a five-stage ring oscillator based on such devices, the latter validating the large-signal prediction capability of the model.

## DISCUSSION

A physics-based large-signal compact model of four-terminal 2D semiconductor-based FETs, implemented in Verilog-A, has been presented. The model captures the terminal charges and capacitances covering all the operation regimes, so accurate time domain simulations and frequency response studies at the circuit level are feasible within the validity of the quasi-static approximation. The model can be incorporated to existing commercial circuit simulators enabling the simulation of digital and RF applications based on emerging 2D technologies. We have checked that the model outcome is consistent with our experimental measurements of MoS$_2$-FET devices targeting RF applications. Finally, a design of a three-stage ring oscillator has been carried out to exhibit the potential of the TCAD tool presented and showing the feasibility of using this technology in switching applications in the gigahertz range.

## METHODS

### Fabrication

The MoS$_2$ FET reported in this paper belongs to the same batch as the devices published in ref. [27] by some of the authors. The fabrication process begins with patterning two embedded gate fingers on intrinsic Si/SiO$_2$ (>20 kΩ cm). Then, the embedded 150-nm gate metal stack consisting of 2/23 nm Ti/Au was defined and deposited by using electron beam lithography (EBL) and e-beam evaporation. Large area atomic single layer CVD MoS$_2$, grown by a standard vapor transfer process as described in ref.,[27] was then transferred by poly(methyl methacrylate)-assisted wet transfer. Phosphoric acid etching was used to connect the embedded gate fingers and the gate pad. The active MoS$_2$ channel was etched using Cl$_2$/O$_2$ plasma. Finally, source and drain (S/D) contacts consisting of 2/70 nm Cr/Au were patterned through a final EBL step.

### Electrical measurements

The electrical DC characterization was done on a Cascade Microtech Summit 11000B-AP probe-station using an Agilent B1500A parameter analyzer. Microwave performance was characterized using an Agilent two-port vector network analyzer (VNA-E8361C). All measurements were taken at room temperature, in ambient atmosphere, and in the dark. The intrinsic microwave performance is obtained after de-embedding the measured data using OPEN and SHORT structures. The OPEN de-embedding was performed on the as-measured device-under-test by etching away the MoS$_2$ in the active regions. The SHORT de-embedding is subsequently carried out by depositing a strip of metal across the channel region, shorting out all pads.

### Circuit simulations

The developed large-signal model of 4T 2DFETs is implemented in Verilog-A and included as a separate module in Keysight© Advanced Design System (ADS), a popular electronic design automation software for RF and microwave applications. It calculates the electrostatics by iteratively solving Eqs. (1) and (3) during run-time using a construct,[21,22] and computes the static and dynamic response of the device by evaluating Eqs. (5) and (6), respectively, using the compact expressions derived in this work.

## DATA AVAILABILITY

The data that support the findings of this study are available from the corresponding author upon reasonable request.

## CODE AVAILABILITY

The large-signal model of 2DFETs implemented in Verilog-A is available from the corresponding author upon reasonable request.

Received: 2 July 2019; Accepted: 13 November 2019;
Published online: 04 December 2019

## ACKNOWLEDGEMENTS

The authors would like to thank the financial support of Spanish Government under projects TEC2017-89955-P (MINECO/AEI/FEDER, UE), TEC2015-67462-C2-1-R (MINECO), and RTI2018-097876-B-C21 (MCIU/AEI/FEDER, UE). F.P. and D.J. also acknowledge the support from the European Union's Horizon 2020 Research and Innovation Program under Grant Agreement No. 785219 GrapheneCore2. A.G. acknowledges the funding by the Consejería de Economía, Conocimiento, Empresas y Universidad de la Junta de Andalucía and European Regional Development Fund (ERDF), ref. SOMM17/6109/UGR. E.G.M. gratefully acknowledges Juan de la Cierva Incorporación IJCI-2017-32297 (MINECO/AEI). A.T.-L. acknowledges the FPU program (FPU16/04043). D.A. acknowledges the Army Research Office for partial support of this work, and the NSF NASCENT ERC and NNCI programs.

# Supplementary Information: Large-signal model of 2DFETs: Compact modeling of terminal charges and intrinsic capacitances

Francisco Pasadas,[1,*] Enrique G. Marín,[2,3] Alejandro Toral-López,[2,4] Francisco G. Ruiz,[2] Andrés Godoy,[2,4] Saungeun Park,[5] Deji Akinwande,[5] and David Jiménez[1]

[1]*Departament d'Enginyeria Electrònica, Universitat Autònoma de Barcelona, 08193, Spain*

[2]*Departamento de Electrónica y Tecnología de Computadores, Universidad de Granada, 18071, Spain*

[3]*Dipartimento di Ingegneria dell'Informazione, Università di Pisa, 56122, Italy*

[4]*Excellence Research Unit "Modeling Nature" (MNat), Universidad de Granada, 18071, Spain*

[5]*Department of Electrical and Computer Engineering, The University of Texas, Austin, TX 78758, USA*

[*]E-mail: francisco.pasadas@uab.es

## Extended transfer characteristics of the MoS$_2$ FET under test

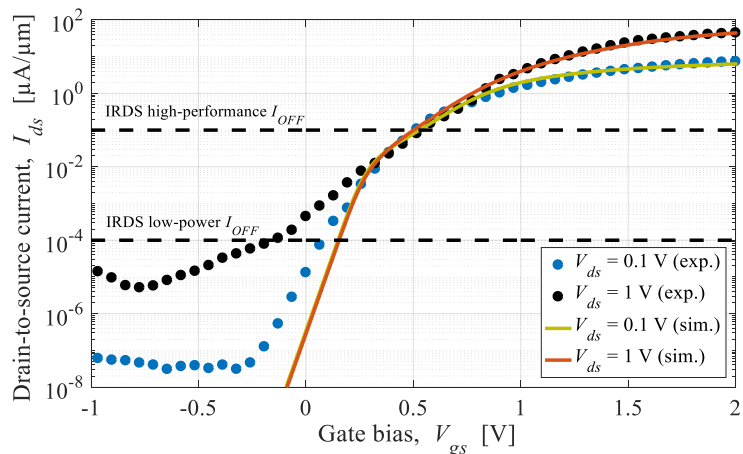

**Fig. S1** – Extended transfer characteristics covering all operation regimes for two different drain voltages. Dashed lines highlight both the IRDS high-performance and low-power off current limits.

Fig. S1 shows the transfer characteristics (TCs) over an extended range of gate voltages. According to our simulations (solid lines), the threshold voltage of the analyzed device is $V_{th}$ ~ 0.42 V. It can be observed that from the near-threshold down to the subthreshold region, the model departs from the experimental DC measurements (symbols). This is likely to happen because of the channel ambipolarity [1],[2], which has not been considered in the model. In spite of the *n*-type contact design, the fabricated device was not optimized to suppress hole conduction. Consequently, a residual hole current at negative $V_{gs}$ appears as can be expected in intrinsic or low doped channels. It is worth noticing that the mismatch between the model and the experimental measurements is observed for current values below the IRDS limit for the OFF current in both low-power ($10^{-4}$ µA/µm) and high-performance applications ($10^{-1}$ µA/µm) (horizontal dashed lines in Fig. S1) what limits its impact in practical circuit applications and has a negligible effect in the circuit-level predictions of the model. Nevertheless, our model works properly for other devices that do not exhibit the residual hole current, as demonstrated in Refs. [3], [4].

## Simplified capacitance model used to compare the proposed model

Equations 13 and 14 of the main manuscript are written here for convenience:

$$Q_{gb} = Q_{2D} \frac{2\left(3C_{tb}\left(e^{u_d}(1+u_s) - e^{u_s}(1+u_d)\right) + e^{u_s}e^{u_d}\left(C_{dq}+C_{tb}\right)\left(u_d^3 - u_s^3\right)\right)}{6C_{tb}\left(e^{u_d} - e^{u_s}\right) + 3e^{u_s}e^{u_d}\left(C_{dq}+C_{tb}\right)\left(u_d^2 - u_s^2\right)}$$

$$Q_d = -Q_{2D}\frac{2}{15}\frac{3u_d^5 - 5u_d^3 u_s^2 + 2u_s^5}{\left(u_d^2 - u_s^2\right)^2}$$

(S.1)

$$C_{ij} = -\frac{\partial Q_i}{\partial V_j} \quad i \neq j \qquad C_{ij} = \frac{\partial Q_i}{\partial V_j} \quad i = j \qquad i,j = g,d,s,b \quad \text{(S.2)}$$

The well-known simplified capacitance model presented for bulk MOSFETs [5] and later used to model 2DFETs [4] is obtained by evaluating the partial derivatives in (S.2) using the following simplified expressions for the terminal charges:

$$Q_{gb} = -(Q_d + Q_s)$$

$$Q_d = Q_{n0}\frac{4 + 8\eta + 12\eta^2 + 6\eta^3}{15(1+\eta)^2}$$

$$Q_s = Q_{n0}\frac{6 + 12\eta + 8\eta^2 + 4\eta^3}{15(1+\eta)^2}$$

$$Q_{n0} = -WL\frac{Q_{net}(x=0) + Q_{net}(x=L)}{2}$$

(S.3)

This simplified charge model uses an empirical function of the drain bias, $\eta$, to provide a continuum between the different regions of the transistor operation [4]–[6] going from the linear regime $\eta = 1$ to the saturation regime $\eta = 0$. It also assumes that the net charge density is linear along the channel ($Q_{n0}$), being this approach accurate only when either a low drain bias is applied or if the device is operated in the subthreshold regime [5], which in turn is in agreement with the comparison shown in FIG. 4 of the main text. In addition, Fig. S2 depicts the charge density profile along the 2D channel for different drain biases showing that it departs from linearity for higher drain biases, a fact that is captured by the model presented in this work (S.1).

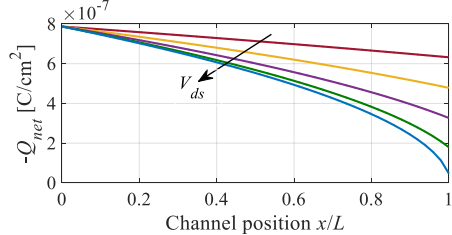

**Fig. S2** – Channel charge density along the normalized channel length of the $n$-type MoS$_2$ based FET described in Table I of the main text ($V_{gs} - V_{g0} = 1$V; $V_{ds} = 0.2$ to 1V in steps of 0.2V).

Note that using the set of equations in (S.1) instead of the set in (S.3), provides a picture more consistent with the underlying physics, as demonstrated in the main text of this work (see Fig. 5 of the main text). Specifically, the difference between both models is shown in solid lines (S.1) and dashed lines (S.3) in Fig. 4 of the main text.

# Five-stage ring oscillator based on bilayer MoS$_2$ transistors

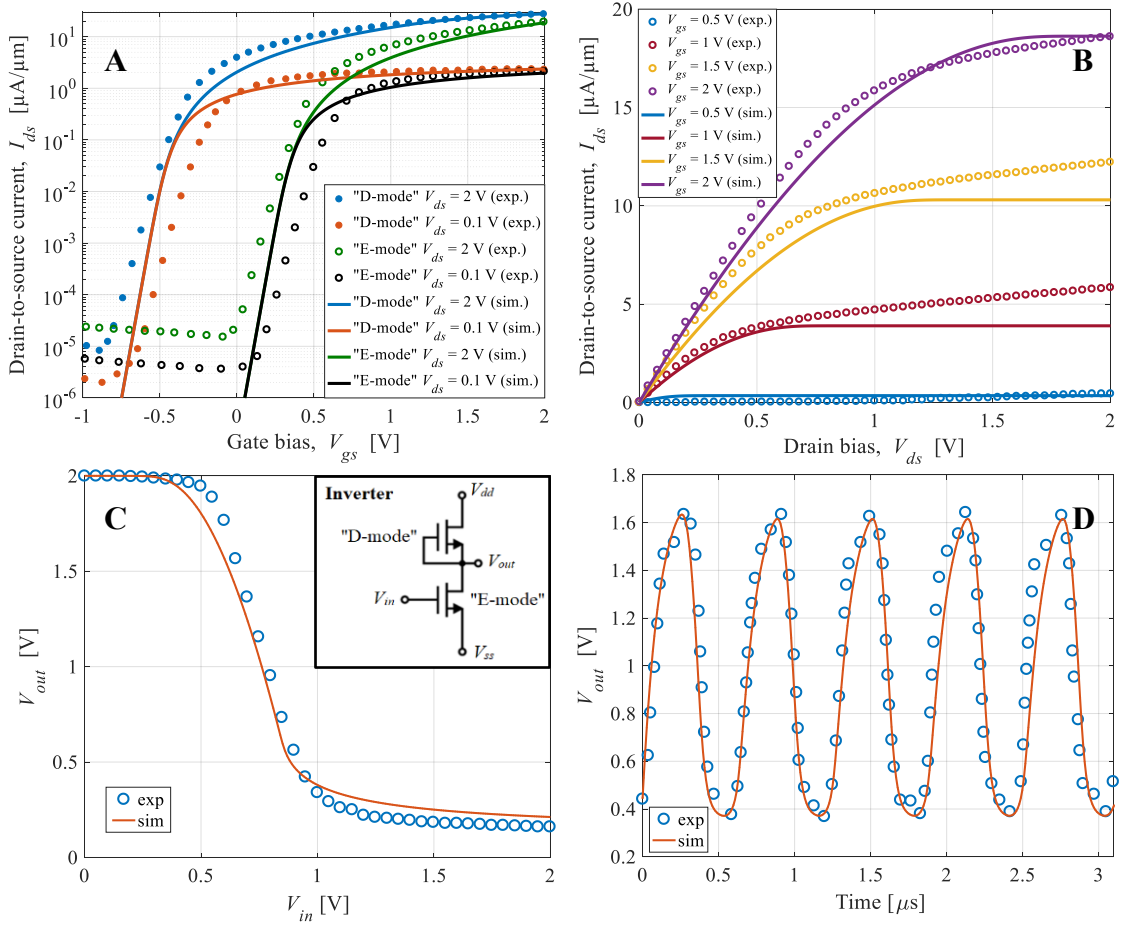

**Fig. S3** – Measurements (symbols) and simulations (solid lines) of **A** the transfer characteristics of depletion (D-) mode and enhancement (E-) mode bilayer MoS$_2$ FETs; **B** the output characteristics of the E-mode bilayer MoS$_2$ FET; **C** the output voltage as a function of the input voltage for a bilayer MoS$_2$ logic inverter (inset: schematics of the bilayer MoS$_2$ based inverter); and **D** the output voltage of a five-stage ring oscillator at a supply bias of 2V. The fundamental oscillation frequency is ~1.6 MHz corresponding to a propagation delay per stage of 62.5 ns.

To assess the validity of the proposed compact model, we have performed a variety of simulations comprising dc transfer (TCs) and output characteristics (OCs), the inverter characteristics and the performance of a five-stage ring oscillator based on depletion (D-) and enhancement (E-) mode bilayer MoS$_2$ FETs reported in Ref. [7]. Both D- and E-mode 2DFETs effectively shift their threshold voltage by about ~0.76V due to a difference in the work function of each metal gate employed in the fabrication process. Specifically, the metals were Al and Pd [7]. Table S1 shows the input parameters used for simulating both devices, where a difference of 0.8V has been applied to the parameter $V_{g0}$ to split the FET operation into D- and E- modes. This parameter has been defined in the main text as a quantity that comprises work-function differences between the gate materials and the 2D channel and any possible additional charges due to impurities or doping.

Fig. S3A shows the comparison between measurements and simulations of the TCs of both devices. Similarly to the TCs of our device (see Fig. S1), residual holes dominate the current at negative gate voltages. It is worth noticing a shift in the threshold voltage likely induced by the drain bias, which is not captured by our model. However, that effect is not apparent in the device analyzed in the main manuscript. Fig. S3B shows the comparison of the OCs of the E-mode bilayer MoS$_2$ FET. In the saturation region, a mismatch between the model prediction and the dc measurements is detected which we attribute to a channel-length modulation effect [5], not considered in our model. In spite of these small mismatches the model captures to a very good degree large-signal operation. In particular, Fig. S3C shows the output voltage as a function of the input voltage in a bilayer MoS$_2$ based inverter formed by the series combination of a D-mode FET in resistor configuration and an E-mode FET (see inset of Fig. S3C). A voltage gain close to 5 is achieved and used to build a five-stage ring oscillator. In doing so, a loop is constructed by connecting five inverters in series. Fig. S3D shows the output voltage switching at a fundamental frequency of ~1.6 MHz corresponding to a propagation delay per stage of 62.5 ns.

TABLE S1. Input parameters of the D- and E-mode bilayer MoS2 FETs reported in [7].

| | | | |
|---|---|---|---|
| $L$ | 1 µm | $\mu_0$ | 22 cm$^2$/Vs |
| $W$ | 4 µm | $m^K$ [8] | 0.542$m_0$ |
| $L_t$ | 18 nm | $m^Q$ [8] | 0.579$m_0$ |
| $L_b$ | 285 nm | $m_0$ | 9.11·10$^{-31}$ kg |
| $\varepsilon_t$ | 25 | $g_K$ | 2 |
| $\varepsilon_b$ | 3.9 | $g_Q$ | 6 |
| $V_{g0}$ | -0.4 V and 0.4 V | $\Delta E_2$ [8] | 0.066 eV |
| $V_{b0}$ | 0 V | $T$ | 300 K |